\documentclass[
 prl,
 reprint,
 floatfix,
 amsmath,amssymb,
 showpacs,
 superscriptaddress,
 citeautoscript,
 nonatbib
]{revtex4-1}

\usepackage{xr}
\makeatletter
\newcommand*{\addFileDependency}[1]{
  \typeout{(#1)}
  \@addtofilelist{#1}
  \IfFileExists{#1}{}{\typeout{No file #1.}}
}
\makeatother

\usepackage{amssymb}

\usepackage[breaklinks]{hyperref}%hyperef package does internal and external linking (labels, urls)
\usepackage[all]{hypcap}%addition to hyperref, it ensures figure is correctly displayed (and not only the caption) when link is clicked
\makeatletter
\renewcommand{\fnum@figure}[1]{\textbf{Fig.~\thefigure~}}
\makeatother

\hypersetup{
    plainpages=false,
    unicode=false,                          % non-Latin characters in AcrobatĂŻÂżÂ˝s bookmarks
    pdfborder={0 0 0},
    pdftoolbar=true,                        % show AcrobatĂŻÂżÂ˝s toolbar?
    pdfmenubar=true,                        % show AcrobatĂŻÂżÂ˝s menu?https://www.overleaf.com/13403665mnpkhvmzmbnjhttps://www.overleaf.com/13403665mnpkhvmzmbnj
    pdffitwindow=false,                     % window fit to page when opened
    pdfstartview={FitH},                    % fits the width of the page to the window
    pdftitle={},                            % title
    pdfauthor={},                           % author
    pdfproducer={LNCMI-CNRS-UGA-UPS-INSA},  % producer of the document
    pdfkeywords={High} {Magnetic} {Fields}, % list of keywords
    pdfnewwindow=true,                      % links in new window
    linktoc=section,
    colorlinks=true,                        % false: boxed links; true: colored links
    linkcolor=blue,                         % color of internal links
    citecolor=red,                          % color of links to bibliography
    filecolor=magenta,                      % color of file links
    urlcolor=blue                           % color of external links
}

\usepackage{graphicx}% Needed to include png/jpg/pdf figure files
\usepackage{float}
\usepackage{amsmath}
\usepackage{bm}
\usepackage{color, soul}
\usepackage[normalem]{ulem}
\usepackage{siunitx}
\usepackage{upgreek}
\usepackage[usenames,dvipsnames]{xcolor}
\usepackage{textcomp}
\usepackage{chemformula}
\usepackage[normalem]{ulem}
\usepackage{rotating}
\begin{document}
\author{Xiangzhou Zhu}
\affiliation{Physics Department, TUM School of Natural Sciences, Technical University of Munich, 85748 Garching, Germany}
\affiliation{Atomistic Modeling Center, Munich Data Science Institute, Technical University of Munich, Garching, Germany}
\affiliation{Munich Center for Machine Learning (MCML), Munich, Germany}

\author{Patrick Rinke}
\email{patrick.rinke@tum.de}
\affiliation{Physics Department, TUM School of Natural Sciences, Technical University of Munich, 85748 Garching, Germany}
\affiliation{Atomistic Modeling Center, Munich Data Science Institute, Technical University of Munich, Garching, Germany}
\affiliation{Munich Center for Machine Learning (MCML), Munich, Germany}

\author{David A. Egger}
\email{david.egger@tum.de}
\affiliation{Physics Department, TUM School of Natural Sciences, Technical University of Munich, 85748 Garching, Germany}
\affiliation{Atomistic Modeling Center, Munich Data Science Institute, Technical University of Munich, Garching, Germany}
\affiliation{Munich Center for Machine Learning (MCML), Munich, Germany}

\title{Predicting the Thermal Behavior of Semiconductor Defects\\with Equivariant Neural Networks}

\date{\today}

\begin{abstract}
The presence of defects strongly influences semiconductor behavior. However, predicting the electronic properties of defective materials at finite temperatures remains computationally expensive even with density functional theory due to the large number of atoms in the simulation cell and the multitude of thermally accessible configurations.
Here, we present a neural network-based framework to investigate the electronic properties of defective semiconductors at finite temperatures efficiently.
We develop an active learning approach that integrates two advanced equivariant graph neural networks: MACE for atomic energies and forces and DeepH-E3 for the electronic Hamiltonian.
Focusing on representative point defects in GaAs, we demonstrate computational accuracy comparable to density functional theory at a fraction of the computational cost, predicting the temperature-dependent band gap of defective GaAs directly from larger scale molecular dynamics trajectories with an accuracy of few tens of meV.
Our results highlight the potential of equivariant neural networks for accurate atomic-scale predictions in complex, dynamically evolving materials.
\end{abstract}

\maketitle

\section{Introduction}\label{sec1}

The electronic structure of point defects critically influences the physical behavior of a wide range of materials, including insulators, metals, and complex oxides, where it governs properties such as ionic transport, magnetism, and catalytic activity.
Understanding these defect states is essential for designing materials with targeted functionalities across energy, quantum, and electronic applications. 
In semiconductor devices specifically, point defects play a crucial role in tailoring electronic and optical properties, impacting performance, reliability, and efficiency\cite{pantelides1978,yu2010,sze2021}.
They can either be introduced to alter the doping level and control charge carrier concentrations or act as unwanted recombination centers that reduce the efficiency of optoelectronic devices such as solar cells \cite{park2018,kirchartz2025}.

Computational modeling of defects in semiconducting materials enables a deeper understanding of their properties and supports the optimization of functional device performance. With quantum mechanical methods like density functional theory (DFT) defect structures, formation energies, charge transition and electronic levels can be predicted \cite{freysoldt2014}.
However, the use of periodic boundary conditions in DFT can create artificial interactions between neighboring defects \cite{puska1998}. 
To mitigate this artificial interaction, electronic-structure calculations of defective semiconductors typically use large supercells containing tens to hundreds of atoms. 
This renders first-principles calculations of defective materials computationally expensive as, e.g., the cost of typical DFT implementations scales as $\mathcal{O}(N^3)$ with the number of atoms, $N$.

Moreover, it is essential to consider thermal vibrations when modeling defect behavior under realistic, ambient conditions\cite{dove1993}. 
Thermal vibrations can influence the geometry and possible migration of defects, further affecting their optical and electronic behavior \cite{youssef2012,glensk2014,qiao2022}. 
For instance, anharmonic vibrations were found to heavily influence the optoelectronic characteristics of point defects and surfaces in halide perovskite semiconductors \cite{cohen2019a,wang2022c,lodeiro2020,delgado2025}.
Capturing such behavior typically requires molecular dynamics (MD) simulations that can sample the ensemble of thermally accessible configurations in the material \cite{gilmer1995,grabowski2009,leyssale2014,hizhnyakov2014,cohen2019a}.
Computing the electronic properties of defective semiconductors -- represented by large supercells in the calculations -- along MD trajectories makes the computational cost even more prohibitive. 

In recent years, the progress of machine learning (ML) techniques created new opportunities for addressing challenges in atomistic modeling of materials in general \cite{butler2018,schmidt2019,Himanen/Geurts/Foster/Rinke:2019,unke2021,li2022e,Kulik/etal:2022} and defect calculations in particular \cite{goryaeva2021,schattauer2022,pols2023,shi2024,mosquera-lois2024, mosquera-lois2025,mohanty2025,ma2025}. 
Among current ML techniques, equivariant message passing neural networks (MPNNs) have proven especially effective for materials modeling. By explicitly incorporating three-dimensional spatial symmetries, they enhance data efficiency, transferability, and predictive accuracy \cite{geiger2022a,batzner2023,corso2024}.   
A flurry of effective MPNNs have been proposed in the development of ML force fields (MLFFs), which can strongly reduce the computational cost in comparison to DFT-based MD calculations while maintaining high accuracy. 
These include NequIP \cite{batzner2022a,batatia2025}, Allegro \cite{musaelian2023}, and MACE \cite{batatia2022a,batatia2022b}. 
These approaches are rapidly gaining traction as compelling alternatives to DFT-based MD simulations, offering defect and surface modeling without the need for explicit first-principles calculations \cite{goryaeva2021,schattauer2022,pols2023,shi2024,mosquera-lois2024, mosquera-lois2025,mohanty2025,ma2025,delgado2025}.

While MPNNs and other ML techniques facilitate the efficient and accurate generation of MD trajectories, they do not intrinsically yield electronic structure information for defective semiconductors, still requiring computationally intensive first-principles methods like DFT. Notably, recent advancements have led to the development of MPNN-based models capable of learning quantum-mechanical Hamiltonians, such as DeepH \cite{li2022c,gong2023,wang2024}, HamGNN \cite{zhong2023,ma2025}, QHNet \cite{yu2023a}, SLEM \cite{zhouyin2025} and MACE-H \cite{MACE-H}.
In Hamiltonian learning, data from first-principles electronic-structure theory such as DFT, are used for training the ML models. Once trained, the ML model predicts the Hamiltonian and with it all electronic properties for a material with a specific atomic structure that is encoded in a set of atomic coordinates. 

Hamiltonian learning has mostly been applied in other contexts of materials modeling, but recent studies highlighted that it can be combined with MLFFs \cite{gu2024a,schwade2025}.
Therefore, Hamiltonian learning may be a promising way to leverage ML methods not only for the MD simulation but also to bypass costly electronic-structure calculations along MD trajectories of defective semiconductors.
For Hamiltonian learning to be effective in this context, a tightly integrated workflow combining MLFFs for MD and Hamiltonian learning for electronic-structure predictions is essential. 
However, to the best of our knowledge, no such workflow currently exists for predicting the properties of defects in semiconductors at finite temperatures.

In this work, we hypothesize that MPNNs, when designed with equivariant architectures, can accurately predict the electronic properties of semiconductor defects under elevated temperatures. This leads to the central research question: Can MPNNs effectively capture temperature-dependent electronic behavior in defective materials, thereby reducing reliance on computationally intensive first-principles methods?

To explore this research question, we unify two state-of-the-art equivariant MPNNs, one for MLFFs and another for modeling quantum Hamiltonians. Specifically, we begin with the MACE-MP0 foundation model \cite{batatia2024}, train MACE MLFFs using DFT data to generate accurate MD trajectories across a range of temperatures, and then investigate how well MACE-based MLFFs reproduce thermally driven atomic dynamics in defective semiconductors compared to DFT-based MD simulations. Using these MLFF-driven MD simulations in combination with further DFT electronic structure calculations, we then train a DeepH-E3 model \cite{gong2023} to learn quantum-mechanical Hamiltonians. To enhance computational efficiency, we propose a workflow embedding both MPNNs within an active learning strategy for MPNN training previously developed for geometry relaxation by one of us \cite{Homm/Laakso/Rinke:2025} and then extended to molecular dynamics \cite{Bhatia/etal:2025}.

We apply our combined formalism to charge-neutral point defects in gallium arsenide (GaAs) to address the following research objectives. Which and how many training configurations do we need? How well does our framework reproduce the DFT electronic structure of point defects in GaAs? And, can we capture the temperature dependence of technologically relevant point defects such as the arsenic antisite (As$_{\mathrm{Ga}}$), which is assumed to have an electronic mid-gap state that significantly influences the electronic behavior of GaAs~\cite{vonbardeleben1986,meyer1987,dabrowski1988}.

\begin{figure*}
        \centering
        \includegraphics[width=1.\linewidth]{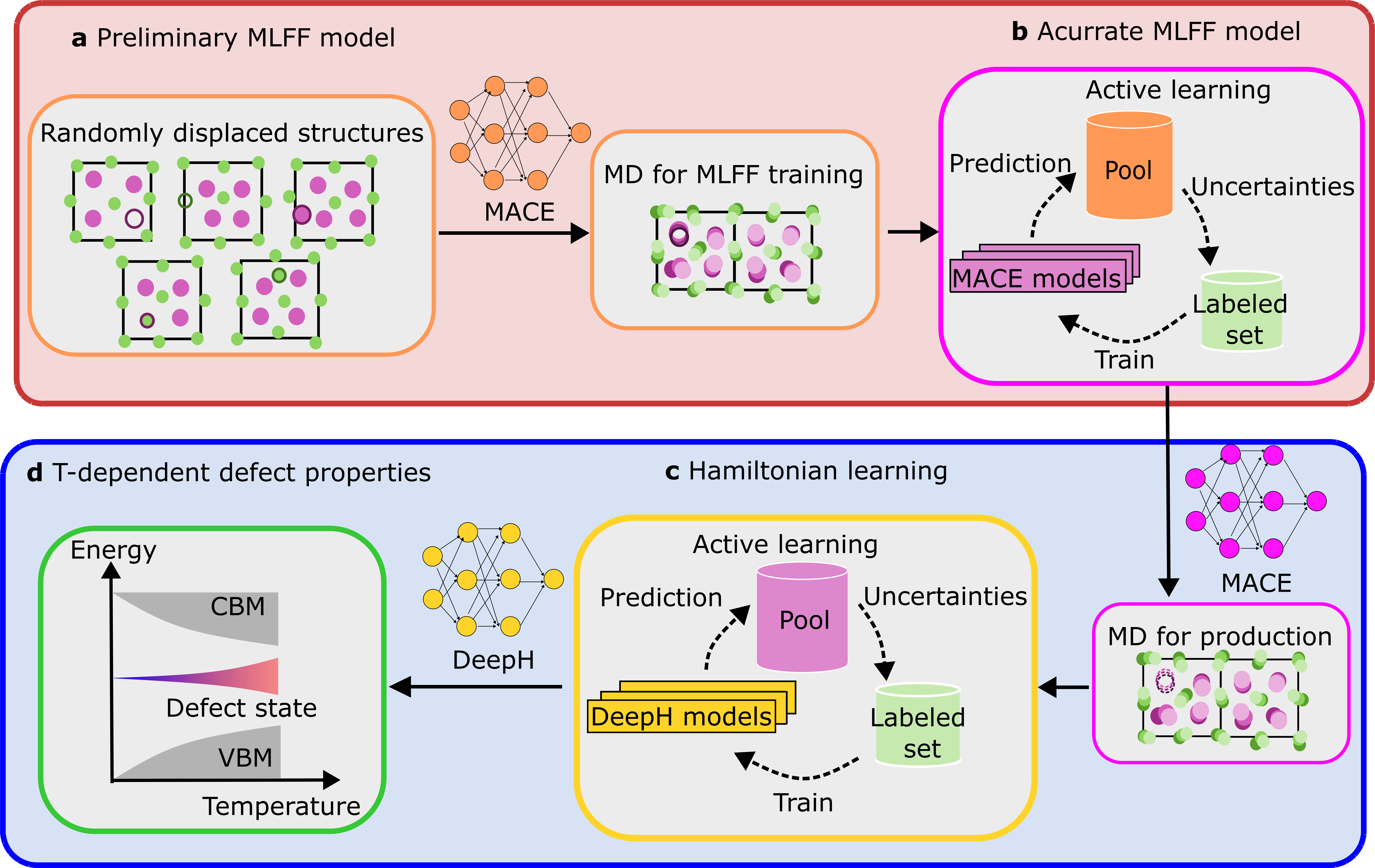}
        \caption{\textbf{MPNN-based workflow for defect predictions.} \textbf{a.} A preliminary MACE-based MLFF is trained on randomly displaced defective structures and used to run short MD simulations for further training. \textbf{b.} An accurate MLFF is obtained via active learning. MD-generated configurations are stored in a pool set, where force uncertainties from an ensemble of MACE models are used to select samples for DFT labeling. These are added to the labeled set for iterative retraining. Longer MD simulations are run and enter the pool set of Hamiltonian learning.
        \textbf{c.} DeepH-E3 models for Hamiltonian learning are also trained in an active learning scheme. Configurations in the pool set, which show high uncertainty in Hamiltonian matrix elements across an ensemble of DeepH models, are selected and ladded to the labeled set for iterative retraining.
        \textbf{d.} The final MLFF and Hamiltonian model are used together for predicting temperature temperature-dependent electronic properties of defects in semiconductors.}
        \label{fig:workflow}
\end{figure*}

    \begin{figure*}
        \centering
        \includegraphics[width=0.7\linewidth]{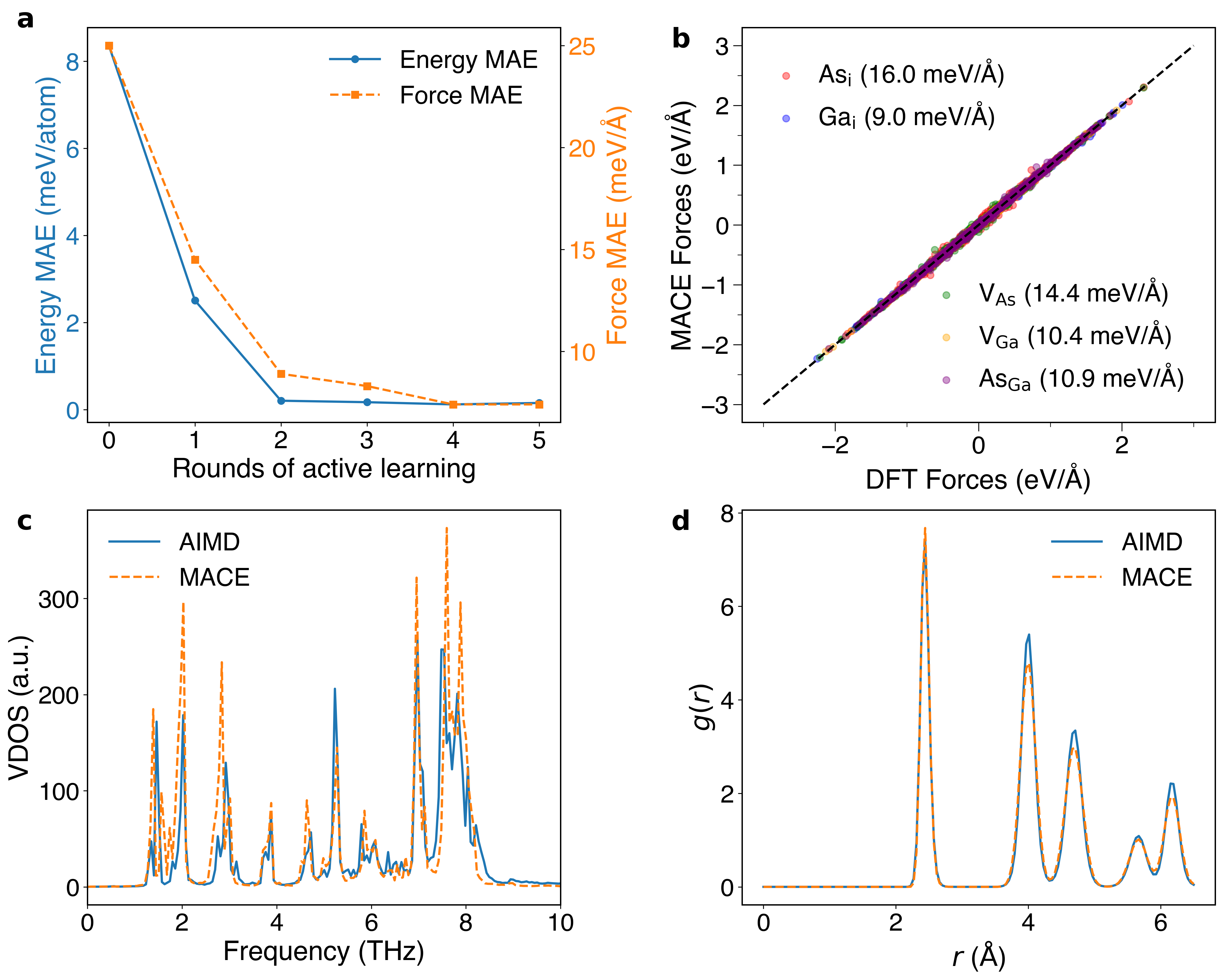}
        \caption{\textbf{Results of the active learning and accuracy of the MACE model in predicting atomic forces and dynamics in defective bulk GaAs.} 
        \textbf{a.} Evolution of the mean absolute error (MAE) in energies (blue solid line) and forces (orange dashed line) during successive rounds of active learning. Zero in the $x$-axis represents the model trained from randomly displaced configurations.
        \textbf{b.} Parity plot of MACE and DFT forces components for five types of defects: As interstitial (As\textsubscript{i}), Ga interstitial (Ga\textsubscript{i}), As vacancy (V\textsubscript{As}), Ga vacancy (V\textsubscript{Ga}), and As antisite (As\textsubscript{Ga}). Each data point represents a force component in a test structure, color-coded by defect type. The dashed line indicates perfect agreement. 
        \textbf{c.} Vibrational density of states (VDOS) and \textbf{d.} radial distribution function, $g(r)$, calculated from AIMD (blue solid line) and MACE-predicted trajectories (orange dashed line), for bulk GaAs with the As\textsubscript{Ga} defect. }
        \label{fig:parity_force}
    \end{figure*}

\section{Results}

\subsection{MPNN-based workflow for defect predictions}

Figure~\ref{fig:workflow} provides an overview of our workflow combining MPNNs for predicting electronic properties of defective semiconductors at elevated temperatures.
It integrates two parts: training a MLFF for dynamical structure predictions and Hamiltonian learning for predicting electronic properties of defective semiconductors.
We first describe our workflow and then apply it to prototypical material systems, namely charge-neutral point defects in bulk GaAs. 
Specifically, we consider five types of defects: the As  (As\textsubscript{i}) and Ga  (Ga\textsubscript{i}) interstitials, the As (V\textsubscript{As}) and Ga  (V\textsubscript{Ga}) vacancies, and the As antisite (As\textsubscript{Ga}).

To accurately sample the structural dynamics of defective GaAs, we employ the MACE MPNN framework \cite{batatia2022a,batatia2022b} 
to train an equivariant MLFF.
Atomic configurations required for MLFF-training could be obtained with ab-initio MD (AIMD), but this is computationally expensive.
Instead, we start from fine-tuning the MACE-MP0 foundation model \cite{batatia2024} by retraining it using the DFT data obtained for 256 randomly displaced configurations of 64-atom defective GaAs supercells (see Figure~\ref{fig:workflow}a and Methods section).
We initially included only four out of the five defect types, and added the As\textsubscript{Ga} defect later when training the MLFF (see below).
We find that this initial, fine-tuned model provides reasonable but not sufficient accuracy compared to DFT results, with forces showing a root mean absolute error (MAE) of approximately 25~meV/\AA~(see Methods section). 

We then run MD calculations at 500~K  
using the initial MLFF for all five defect types, generate 500 configurations of each defect type with a 128-atom supercell, and train an improved MACE model as the next step using an active learning strategy following the work of Homm \textit{et al.} \cite{Homm/Laakso/Rinke:2025} and Bhatia \textit{et al.} \cite{Bhatia/etal:2025} (see Figure~\ref{fig:workflow}b).
Here, a relatively high temperature of 500~K is chosen to explore a broader region of configuration space. 
We generate a \textit{pool of structures} consisting of MD-generated structures and then predict the forces acting on each atom using three different MACE models trained with different random seeds.
We then calculate the force uncertainties by comparing the variations in predicted forces across the models. 
Configurations with force uncertainties exceeding a threshold of 10~meV/\AA, 
which is a commonly used force-convergence criterion in DFT geometry optimizations, are selected and added to the \textit{labeled set} for training new MACE models with DFT data.
This threshold ensures that only those configurations for which force predictions are not yet reliable are included.
The active learning loop can be repeated until a sufficiently accurate model is obtained.
Once the training procedure is concluded, we are in a position to carry out long MD production runs for five defect types in GaAs with the final model in LAMMPS \cite{thompson2022}. 
This provides sets of structures with thermal fluctuations for analyzing the electronic properties of defects at finite temperatures.

To accurately predict electronic properties of the different defective supercells including their thermally-sampled atomic configurations, we employ Hamiltonian learning (see Figure~\ref{fig:workflow}c).
Performing DFT calculations on every MD snapshot for predicting electronic properties is time-consuming, especially for large system sizes of defective systems.
We overcome this hurdle via the DeepH model to bypass costly DFT calculations.
To enhance temperature transferability, we generate our training sets with the final MACE MLFF in MD calculations at 100~K and 500~K, for each of the five defective structures, and employ a k-means clustering technique to select structures (see Methods section).

Specifically, we train a DeepH model \cite{li2022c,gong2023,wang2024} 
using the same active learning scheme as before to enhance accuracy and to efficiently explore configuration space (see Figure~\ref{fig:workflow}c). 
First, we initialize the active learning procedure by selecting 100 structures from MD simulations of the five defects at 100~K and 500~K. These two temperatures are chosen to include structures that are representative for both the low and high temperature-regime. 
Three independent DeepH models are then trained on this initial labeled dataset. 
Next, like above for training the MLFF, we collect 1000 unlabeled atomic configurations uniformly sampled in time from the MD simulations to form a \textit{pool of structures} (see Methods section). 
Finally, we estimate the uncertainty across the three models by calculating the maximum of standard deviations across the Hamiltonian matrices predicted by the models, which is more sensitive to rare outliers.
Unlike for the MLFFs, where we applied a fixed uncertainty threshold, we here add the 30 structures with highest uncertainty into the \textit{labeled set} for training DeepH models with DFT data.
However, to avoid overfitting to a particular defect type, we add an equal number of samples for each defect type, at both temperatures, to the training set  (see Methods section).

We iterate the active learning cycle until the accuracy of eigenvalues near the band gap reaches saturation, indicating no further improvement. This yields a DeepH model capable of predicting Hamiltonians for defective systems, which we subsequently evaluate in terms of its performance and transferability across varying temperatures, system sizes, and defect types.
An accurate model will allow us to predict electronic properties of defective semiconductors at elevated temperatures (see Figure~\ref{fig:workflow}d).
To evaluate the model performance, we construct a test set consisting of 75 unseen configurations sampled from independent 5~ps MD simulations at 100~K, 300~K, and 500~K for each defect type.

\begin{figure*}
    \centering
    \includegraphics[width=\linewidth]{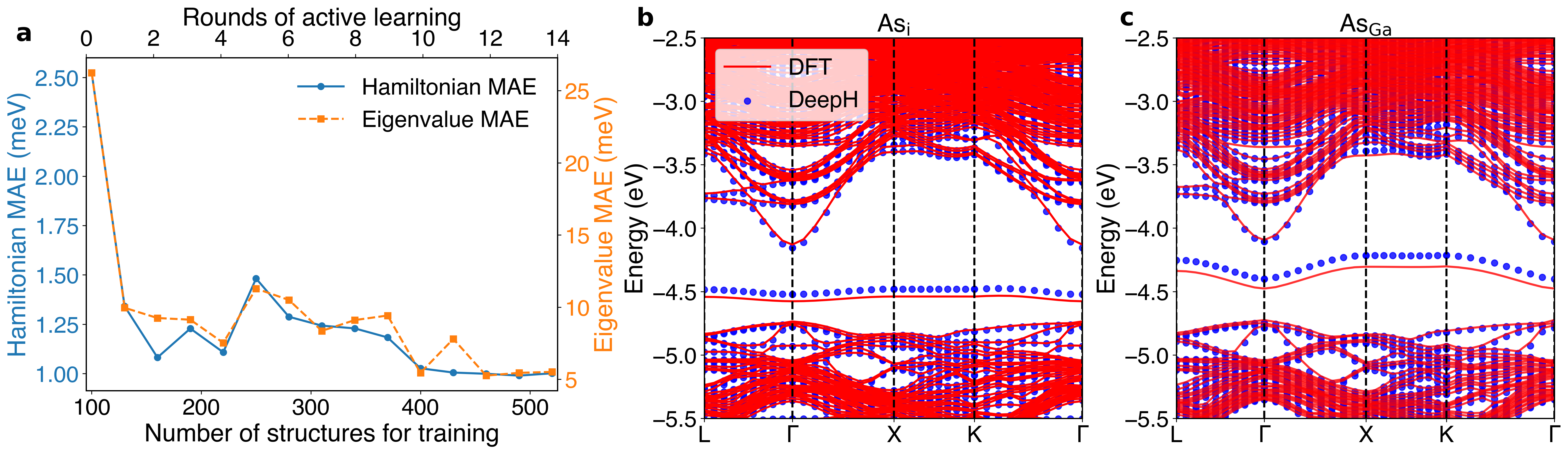}
    \caption{\textbf{Active learning and benchmarking of DeepH model for predicting the electronic structure in defective GaAs.} 
     \textbf{a.} Mean absolute error (MAE) of Hamiltonian matrix elements (blue solid line) and electronic eigenvalues for 30 bands around the band gap (orange dashed line) predicted by DeepH in comparison to DFT as a function of training set size.
    \textbf{b, c.} Electronic band structures of the neutral point defects As\textsubscript{i} (panel b) and As\textsubscript{Ga} (panel c) obtained from DFT (red solid lines) and the DeepH model (blue dots). The data were obtained considering a random snapshot taken from a 432-atom supercell MD simulation with the MACE MLFF at 100 K.}
    \label{fig:deeph-band}
\end{figure*}

\subsection{Prediction of thermal atomic motions in defective GaAs and benchmark of the MLFF}

Using the procedure described above (see Figure~\ref{fig:workflow}), we performed several active-learning rounds to train MACE models capable of predicting energies and forces within a single MLFF for five types of defects in GaAs. 
As shown in Figure~\ref{fig:parity_force}a, the active-learning process converges rapidly, as evidenced by the decreasing mean absolute error (MAE) in energies and forces between DFT and MLFF predictions for a test set of structures not used during training. 
We selected the MACE model obtained after round 2 of the active-learning loop, as the force predictions had already reached high accuracy, with an MAE below 10~meV/\AA.

Figure~\ref{fig:parity_force}b shows a comparison of atomic forces predicted by the final MACE model and obtained in DFT. 
The root mean square errors (RMSEs) range from 9~~meV/\AA (Ga interstitial, Ga\textsubscript{i}) to 16~~meV/\AA (As interstitial, As\textsubscript{i}) and indicate that the model is accurate.
{In the Supplemental Material, we present additional results at 100~K and 500~K.}

To further assess the reliability of the MACE MLFF, we present dynamical observables for the representative case of an arsenic antisite defect in GaAs and benchmark the results against those obtained from DFT-based MD simulations at 300 K (see  Figure~\ref{fig:parity_force}). 
Figure~\ref{fig:parity_force}c shows the vibrational density of states (VDOS) and Figure~\ref{fig:parity_force}d the radial distribution function (RDF), $g(r)$.
The agreement of VDOS and RDF calculated with the MACE MLFF in comparison to AIMD confirms that the trained model accurately captures the structural dynamics in defective GaAs.
Similar findings for the other four types of defects in bulk GaAs are shown in the Supplemental Material.
Therefore, the MLFF created in our active learning procedure successfully predicts thermal atomic motions for five types of defects in bulk GaAs at various temperatures.
We use the MACE MLFF for computing finite temperature configurations and proceed with predicting electronic properties via Hamiltonian learning.

    \subsection{Prediction of Hamiltonian and electronic structure of defective GaAs with DeepH model}

\begin{figure*}
        \centering
        \includegraphics[width=0.7\linewidth]{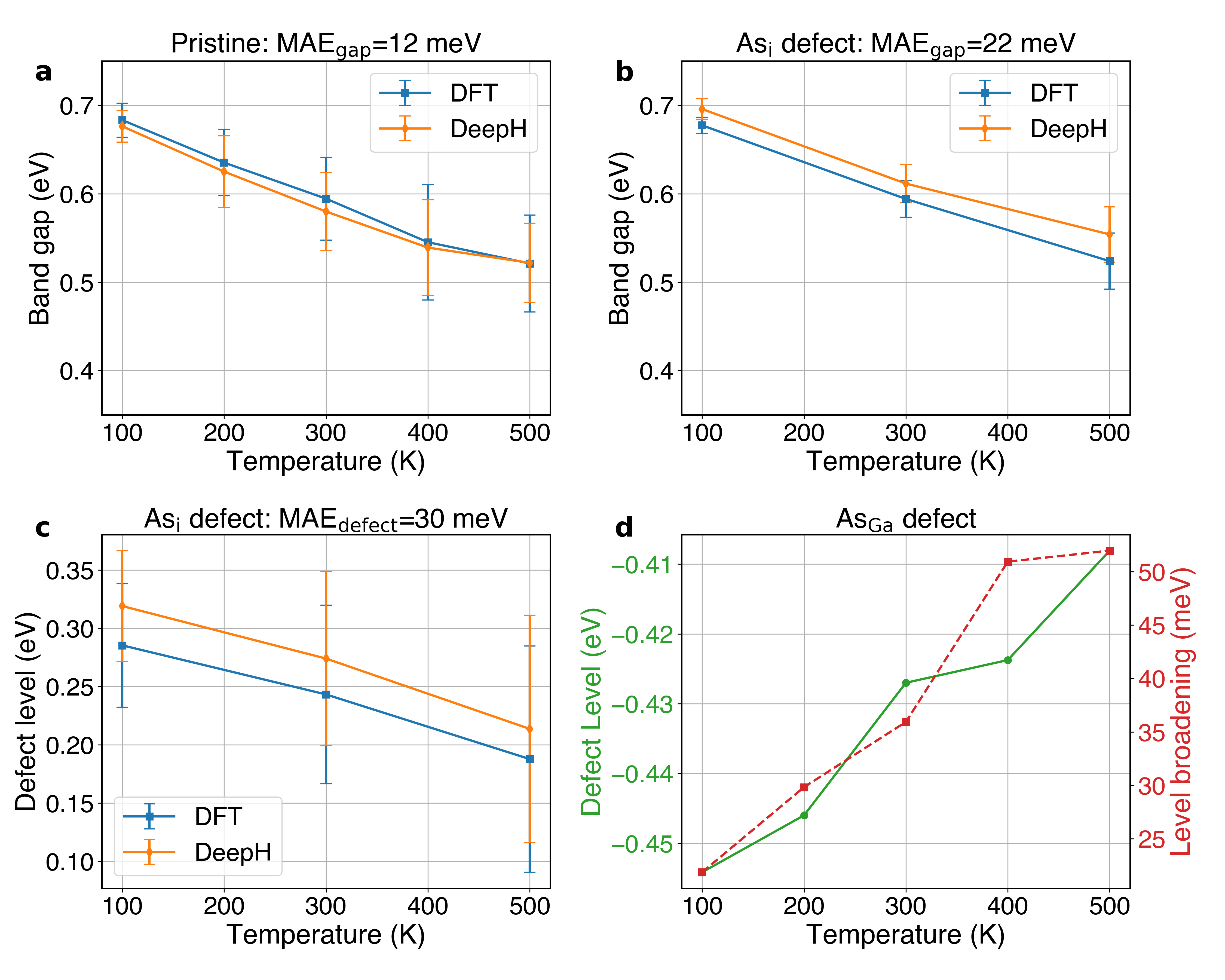}
        \caption{\textbf{Temperature-dependent electronic properties of pristine and defective GaAs.} \textbf{a, b, c.} Band gap of pristine GaAs (panel a) as well as  as a function of temperature, \textbf{b.} Band gap of GaAs containing As\textsubscript{i} and  \textbf{c.} defect level energy, reported  with respect to the valence band maximum, of As\textsubscript{i} in GaAs as a function of temperature, computed by DFT (blue) and DeepH (orange). \textbf{d.} DeepH predicted defect level energy and broadening as a function of temperatures, where the conduction band minimum (CBM) is set to zero. The level broadening is defined as the standard deviation of the defect levels.}
        \label{fig:deeph-pris}
    \end{figure*}

Using our active-learning procedure (see Figure~\ref{fig:workflow}), we completed a number of rounds to train DeepH models for predicting the Hamiltonian and energy eigenvalues within one model for five types of defects.
The accuracy of the predicted Hamiltonian and its eigenvalues in comparison to DFT data are shown across each round in Figure~\ref{fig:deeph-band}a.
We find that the Hamiltonian matrix elements converge to an MAE of approximately 1~meV and the energy eigenvalues to an MAE of about 10~meV after 12 active-learning rounds, using a total of 460 training structures. Adding additional data does not further reduce the MAE. Based on these results, we therefore employ the model trained on 460 structures to predict, benchmark, and investigate the electronic properties of defective GaAs using DeepH.

In Figure~\ref{fig:deeph-band}b-c, we compare electronic band structures calculated from Hamiltonians obtained in DFT and our DeepH model. 
In this comparison, we use a 432-atom supercell and select two types of neutral point defects, As\textsubscript{i} and As\textsubscript{Ga}, to sample their structure at 100~K in MD calculations with our MLFF. 
Note that the supercell we used in the training set contained only 128 atoms.
The DFT and DeepH results agree well for the two cases here and further ones reported in the Supplemental Material. 
Specifically, for most bands the error is below 30~meV, although we observe that the defect level in 432-atom supercell can deviate by up to 50~meV.
Even though errors in the range of 20–50~meV are not negligible, they should be interpreted in the context of typical uncertainties inherent in DFT calculations of solids, as these calculations form the foundation of the training data.
These findings demonstrate high accuracy of our approach with transferability across system size, defect types, and temperatures.

In the electronic band structures (see Figure~\ref{fig:deeph-band}b-c), the defect levels are seen as relatively flat bands within the band gap: 
the As\textsubscript{i} defect level lies approximately 0.3 eV above the valence band maximum (VBM) and the As\textsubscript{Ga} defect appears mid-gap. 
A certain dispersion of the electronic states derived from the defects is visible: it is more pronounced for the mid-gap state associated with As\textsubscript{Ga}, whereas the As\textsubscript{i} state -- more tightly confined around the interstitial site (see below) -- shows a weaker slope. 
This dispersion arises from artificial defect–defect interactions\cite{freysoldt2014}, a well-known artifact of periodic boundary conditions present in DFT and in electronic structure methods more broadly. 
While these interactions remains in finite supercell,  we stress that the efficiency of Hamiltonian learning enables simulations of much larger systems at modest computational cost, where these artificial interactions can be systematically reduced (see Supplemental Material).

Importantly, DeepH reproduces the dispersion of defect states as calculated in DFT, including its change with supercell size. 
This finding indicates that the model can capture both the short-range chemical effects, responsible for the position of the defect energy level, and the longer-range effects that give rise to the artificial behavior, as encoded in the training data with limited supercell size. 
Once trained, DeepH achieves a more than 30 times speedup over DFT for a 432-atom supercell, measured on a node with two Intel Xeon Platinum 8168 CPUs (see Supplemental Material). 
This computational speedup enables efficient prediction of electronic properties in defective systems across thousands of MD snapshots.

\subsection{Prediction and analysis of temperature-dependent electronic properties of defective GaAs}
Since DeepH accurately predicts the electronic structure of thermally disordered, defect-containing configurations, we now turn to analyzing temperature-dependent electronic properties.
Specifically, we compute statistics of the band gap and the defect level over ensembles of MD snapshots, enabling us to extract their thermal averages and distributions in comparison to DFT.

First, we calculate the band gap of pristine GaAs as a function of temperature with both DFT and DeepH (see Figure~\ref{fig:deeph-pris}a). 
Both methods predict a monotonic decrease in band gap as temperature increases, reflecting the well-known band gap renormalization due to lattice vibrations. 
Notably, the DeepH model was trained exclusively on defective structures, yet it generalizes well to the pristine system. 
The MAE of 12~meV indicates good transferability and remains small relative to the typical numerical errors across DFT implementations for gapped bulk materials. Figure~\ref{fig:deeph-pris}b presents the corresponding results for the As\textsubscript{i} interstitial defect.
Again, our workflow with DeepH accurately reproduces the DFT band gap across temperatures, with an MAE of 22~meV. Figure~\ref{fig:deeph-pris}c presents the defect level energy reported with respect to the VBM, for which we find a MAE of 30~meV. In passing, we note that the temperature dependence of the band gap is largely unaffected by the presence of As\textsubscript{i}.

Finally, we leverage the DeepH model to predict the temperature dependence of the  As\textsubscript{Ga} defect level.
Among the five defect types considered in our work, As\textsubscript{Ga} is selected here as a representative example because it has been extensively studied and is known to strongly impact the electronic structure due to its mid-gap character \cite{vonbardeleben1986,meyer1987,dabrowski1988}.
We report the defect level energy relative to the CBM and again employ a 250-atom supercell.
As shown in Figure~\ref{fig:deeph-pris}d, the energy difference between defect level and CBM decreases as temperature increases.
This upward shift of the defect level is primarily due to the renormalization of the band edges, rather than intrinsic thermally-induced changes in the energy of the defect state itself.
In addition to the ensemble average of defect energies, we also calculate the thermal fluctuations and quantify the broadening of defect levels as shown in Figure~\ref{fig:deeph-pris}d.
As temperature increases, the defect level not only shifts relative to the CBM but also broadens significantly, indicating their coupling to thermal vibrations.
These findings provide a solid basis for direct comparison with experimental techniques, such as temperature-dependent photoluminescence.
In the Supplemental Material, we show additional results about pressure-dependent electronic properties of defective GaAs, which are found to be in qualitative agreement with experimental data.

\section{Discussion}

Point defects strongly influence the physical behavior of materials.
In semiconductors, they are especially critical, as they directly affect device performance in applications such as photovoltaics.
Therefore, a detailed understanding and characterization of defect properties is essential.
While first-principles methods such as DFT provide valuable atomic-scale insights for defective materials, they are computationally costly due to the need to model large supercells and account for a wide range of thermally accessible configurations.
ML models based on MPNNs have greatly accelerated separate computations of atomic dynamics and electronic structure.
Yet, no unified framework currently exists to predict temperature-dependent electronic properties of defective semiconductors in a fully integrated manner.

In this work, we have presented a framework that integrates two MPNNs within an active learning architecture to predict electronic properties of defective semiconductors in a computationally efficient manner.
Specifically, we combine MACE for MLFF training and DeepH for Hamiltonian learning and apply our workflow to predict the structural dynamical and electronic properties of bulk GaAs with point defects.
Our approach combining MPNNs and active learning was demonstrated to synergistically provide accurate predictions of atomic dynamics and electronic properties. 
Standard DFT methods remain costly for large ensembles of configurations required to capture temperature-dependent properties, while ML methods that do not preserve fundamental physical symmetries may lack generalizability across temperatures and defect types, especially when learning Hamiltonians.

In contrast, our framework unifying MACE and DeepH models leverages equivariant graph neural networks to inherently preserve physical symmetries, resulting in transferability and accuracy across temperature regimes and defect structures. 
Our active learning approach further reduces computational efforts by identifying and performing DFT only on the most informative configurations.
We investigated the performance of our approach on well-known defects in the prototypical semiconductor GaAs and found very good accuracy for a number of observables. 
For instance, our method predicts the temperature-dependent band gap of defective GaAs with errors of less than 30~meV, which is small compared to numerical errors that are inherent in DFT implementations of solids.

Finally, we used our approach to predict energy level shifts and level broadenings of As\textsubscript{Ga} -- the EL2 defect state in GaAs -- to not only exemplify the great promise of efficiently exploring defect phenomena but also highlight the critical impact of thermal vibrations on electronic properties.
Altogether, our combined framework holds great promise for efficiently exploring the influence of temperature on the electronic properties in defective semiconducting systems, e.g., when studying other key properties such as  carrier mobility or analyzing the consequences of defect complexes in materials.

Despite these advantages, certain limitations remain to be addressed in future work.
First, the current model only considers neutral point defects, which restricts its applicability especially regarding systems where charged defects are important.
Additionally, the chosen uncertainty metric (maximum standard deviation of Hamiltonians across models) may benefit from further refinement to enhance the performance of electronic observable predictions.
Moreover, we have considered a relatively simple material system as a prototypical semiconductor with well-characterized point defects, and it will be important to investigate the accuracy of MPNN-based models in more complex materials.
Finally, Hamiltonian learning in our study was performed on the basis of PBE training data, but semilocal DFT functionals such as PBE are known to severely underestimate the the band gap. 
Likewise, we employed a relatively small basis set when learning the Hamiltonian, and how well the methods performs for larger basis set sizes can be investigated in future studies.
Integrating higher levels of theory in the training data, such as Hamiltonians computed with hybrid functionals, remains as an important future avenue to improve the prediction accuracy comparable to experiments. 

\section{Conclusions}

In conclusion, this study set out to investigate whether MPNNs can learn the atomic and electronic structure of point defects in semiconductors. 
By combining the MACE MLFF with DeepH Hamiltonian learning in an active learning framework, we found that we can accurately reproduce the electronic structure of point defects in GaAs along a MD trajectories with a moderate amount of DFT training calculations. Our model then facilitated the analysis of the temperature dependence of the electronic structure in the arsenic antisite in GaAs, a defect of technological relevance. These findings contribute to our understanding of point defects in semiconductors and have potential applications in optoelectronic devices such as solar cells. Further research could explore charged defects and polarons, which also affect the device behavior.
 
\section{Methods}
\subsection{DFT dataset preparation}
To construct the training set for the MACE model, we employed a 64-atom supercell for the preliminary model and a 128-atom supercell for the production-level model. 
DFT forces and energies were calculated with the Vienna ab initio simulation package (VASP) \cite{kresse1996}. 
An energy cutoff of 400 eV, energy convergence threshold of $10^{-6}$ and $\Gamma$-centered k-grid of 4$\times$4$\times$4 (2$\times$2$\times$2) were used for 64 (128)-atom supercells.
Ga\_d (3d$^{10}$4s$^{2}$4p$^1$) and As\_d (3d$^{10}$4s$^{2}$4p$^3$) were used as projector-augmented wave (PAW) potentials \cite{blochl1994}. For exchange and correlation, we employed the Perdew-Burke-Ernzerhof (PBE) functional \cite{perdew1996}.

To train the DeepH model, we employed the OpenMX package~\cite{boker2011}, utilizing pseudo-atomic localized basis sets~\cite{ozaki2004} to compute DFT Hamiltonians. For simulations involving 128-atom supercells, we applied an energy cutoff of 300~Ry, an energy convergence threshold of 
$10^{-7}$ Hartree, and a 2$\times$2$\times$2 k-point grid. The pseudo-atomic orbitals Ga7.0-s3p2d2 and As7.0-s3p2d2 were used for gallium and arsenic, respectively~\cite{morrison1993}.

\subsection{Details of MACE training}
The preliminary MACE models \cite{batatia2022a,batatia2022b} were naively tuned on the MACE-MP-0b foundation model \cite{batatia2024} with a training set of randomly displaced 64-atom supercell of 4 types of defects: As interstitial (As\textsubscript{i}), Ga interstitial (Ga\textsubscript{i}), As vacancy (V\textsubscript{As}), Ga vacancy (V\textsubscript{Ga}), each 64 configurations. 
Then, MD simulations were performed with a 128-atom supercell from which 100 structures were selected as an initial training set for training three MACE models. 
The uncertainty was quantified as the maximum standard deviation of atomic forces in the ensemble of the three MACE models.
Configurations showing an uncertainty larger than 0.01 eV/\AA~ during a 5-ps MD run at 500~K were selected and used to retrain the model in two rounds.
In these two rounds of active learning, a new defect type, arsenic antisite (As\textsubscript{Ga}), was introduced by including structures from it generated with MD in the training set. 
Thus, the final training dataset includes five distinct defect types.
To benchmark the resulting MLFF, we perform MD production runs for each type of defect with the MACE MLFF for 30~ps at 300~K. From each trajectory, 10 snapshots were randomly selected, for which DFT forces were computed and compared to the MLFF predictions. 
The RMSE of the forces was then evaluated for these snapshots for all five defect types to assess the accuracy of both the preliminary and final MACE models.

\subsection{Details of DeepH training}
For the DeepH-E3 model \cite{gong2023}, we used the maximum angular momentum in spherical harmonics of 4 to encode geometry information. The node and edge feature vectors in the intermediate layers were constructed from the irreducible representations \texttt{64x0e+32x1o+16x2e+8x3o+8x4e}, and three message passing layers were used.
The DeepH model was initially trained on 100 structures (33 structures as validation) with a training set of 128-atom supercell of 5 types of defects: As interstitial (As\textsubscript{i}), Ga interstitial (Ga\textsubscript{i}), As vacancy (V\textsubscript{As}), Ga vacancy (V\textsubscript{Ga}), As antisite (As\textsubscript{Ga}).
For each type of defect, 10 configurations are selected from 10 ps MD simulations at 100~K and 500~K at different clusters through the k-means clustering technique.
Then, 1000 configurations at 100~K and 500~K are collected in 20 ps (200 configurations for each type of defect) as candidate structures to be further picked in the active learning scheme.
Again, we quantified the uncertainty through the disagreement of the ensemble of three DeepH models, where we estimate it by calculating the root mean square deviation (RMSD) of the Hamiltonian matrix.
To ensure the balance of representative training set, we introduced a constraint in the selection: three configurations were selected for each combination of defect type and temperature. 
As a result, in each round of active learning, 30 structures are selected (5 defect types × 2 temperatures × 3 configurations).
The test set was selected from 5 ps MD simulations at 100~K, 300~K and 500~K that were not seen by the training part.
The error in eigenvalue was calculated by comparing the DeepH prediction and OpenMX calculation.
Specifically, we calculate the MAE of Hamiltonian matrix elements by averaging their absolute deviation, that is $|H_{\text{DeepH}}-H_{\text{DFT}}|$, considering all matrix elements.
Furthermore, we calculate the MAE of eigenvalues by averaging their absolute deviation, that is $|E_{\text{DeepH}}-E_{\text{DFT}}|$, including the 30 bands closest to the band gap. 
For each of these bands, we sampled the electronic eigenvalues along a high-symmetry path L-$\Gamma$-X-K in the Brillouin zone, using a sample of 15 $k$-points between each of the high-symmetry points for 30 bands around band gaps.

\subsection{Band gap and defect level calculations}
The temperature-dependent band gaps and defect levels were calculated by taking an average over 100 snapshots from the MD simulation at specific temperatures.
The band gap of the As\textsubscript{i} defect was calculated with a 128-atom supercell, while to remove the artificial dispersion of the defect levels, they were calculated with a 250-atom supercell.
The MD trajectory is generated from LAMMPS \cite{thompson2022} using the NVT ensemble with a time step of 1 fs  with the MACE force field.

\section{Data availability}

The raw data for all calculations are available on Zenodo (\url{https://doi.org/10.5281/zenodo.17670347}).

%\printbibliography
\bibliographystyle{apsrev4-1}
%%\bibliography{z_revtex_template}% Produces the bibliography via BibTeX.
\bibliography{hamiltonian2025}

\section{Acknowledgments}

Funding provided by Germany's Excellence Strategy (EXC 2089/1-390776260) is gratefully acknowledged.
The authors further acknowledge the Gauss Centre for Supercomputing e.V. for funding this project by providing computing time through the John von Neumann Institute for Computing on the GCS Supercomputer JUWELS at Jülich Supercomputing Centre.

\section{Author contributions}

D.A.E. and P.R. designed and supervised the research. 
X.Z. performed the calculations, benchmarked the parameters, and finalized the workflow with optimized settings.
All authors contributed to the discussions and provided feedback on the manuscript.

\section{Competing interests}

There are no competing interests.

\end{document}